\documentstyle[12pt]{article}

\begin{document}
\thispagestyle{empty}
\rightline{SU-4240-583}
\rightline{June 1994}
\rightline{hep-ph/9407270}
\vskip 2cm
\centerline{{\bf PREDICTIONS OF NONCOMMUTATIVE SPACE-TIME}}
\vglue 1.0cm
\centerline{NGUYEN AI VIET }
\baselineskip=14pt
\centerline{{\it Physics Department, Syracuse University}}
\baselineskip=14pt
\centerline{{\it Syracuse, NY 13244-1130, USA}}
\begin{abstract}
An unified structure of noncommutative (NC) space-time for both gravity and
particle physics is presented. This gives possibilities of testing the idea
of NC space-time at the currently available energy scale. There are several
arguments indicating NC space-time is visible already at the electroweak scale.
This NC space-time predicts the top quark mass $m_t \sim 172GeV$, the Higgs
mass $M_H \sim 241 GeV$ and the existence of a vector meson and a scalar, which
interact universally with the Matter.
\end{abstract}
\smallskip
\newpage
\setcounter{page}{1} \pagestyle{plain}

%
%
{\bf\noindent 1. Introduction}
\vglue 0.4cm
Nowadays, there exist many intelligent theories, which we can neither confirm
nor deny by the currently available experimental data. While using them as a
tool to construct physical models, we always have an uncomfortable doubt:
whether they are more fundamental than a single mathematical tool?

Noncommutative geometry (NCG) has been announced [1] as a new concept of
space-time beyond the ordinary one. Its enormous and increasing influence on
the whole mathematical physics suggests that NCG may have some relevance to
physics. However, its achievements in particle physics [2] have not been enough
to make phenomenologists believe that this sophisticated machinery contains
more than putting known things together. If we want to consider more seriously
the idea of noncommutative (NC) space-time and the 'dogma':
{\it NCG is the geometry of the realistic space-time, that becomes visible at
some high energy scale}, we will have to face the following questions:
{\it i) How can we determine that energy scale ? ii) Can we test the existence
of NC space-time by the currently available facilities ?}
Theoretical predictions for too high energy scales would not convince an
understandably impatient phenomenologist any more. Fortunately, hereafter we
are able to offer some
testable implications of the concept of NC space-time. The starting point is
based on the idea [3] to fix the structure of NC space-time by a
reasonable requirement: the physical space-time must be the same for both
particle physics and general relativity. It turns out that NC space-time has
a K\"ahler metric structure and imposes very strict constraints on the
physical quantities. In fact, these constraints are valid only on a certain
energy scale, because the parameters of the theory evolve independently by
the quantum corrections. However, by renormalisation group arguments, we can
use those constraints as initial values to determine that energy scale and
derive some predictions for the currently available scale.

\vglue 0.6cm
{\bf\noindent 2. Noncommutative Space-time}
\vglue 0.4cm
NCG is given by the triplet $(~{\cal A}~,~ D ~,~
{\cal H}~ )$, where i) the algebra ${\cal A}$, which is not necessarily
commutative, generalizes the algebra $ C^\infty ({\cal M})$ of
continuous functions on the manifold ${\cal M}$ of the ordinary geometry, ii)
the Dirac operator $ D $, which generalizes the exterior derivative $ d $,
satisfies De Rham's theorem $D^2~=~0$ and
iii) the Hilbert space ${\cal H}$ of the fermionic sector.
The NC space-time considered in this paper is the two-copied one based on the
algebra $C^\infty({\cal M})\oplus C^\infty({\cal M})$ [1-6].
A differential geometric structure of this NC space-time has been constructed
by Coquereaux et al [4]. A graded structure including both even and odd
differential elements has been introduced to generate a quartic Higgs potential
in gauge theories. On the other hand, a theory of gravity in NCG has been also
constructed using only anticommutative differential elements [5,6] in
parallelism with the ordinary Einstein theory. As the space-time locally cannot
be different in particle physics and in general relativity, we proposed [3] to
incorporate both structures into an unique one by
introducing a pair of conjungate 'discrete' differential elements instead of a
real one. This is an analogue of the Newman-Penrose formalism of GR. In the
Standard Model this NCG gives the same physical content as in [2,4], provided
it admits a K\"ahler metric. Let us briefly summarize the idea.

The Hilbert space of the fermion sector ${\cal H}$ is the direct sum of the
Hilbert spaces ${\cal H_L}$ and ${\cal H_R}$ of the left- and right-handed
particles: $ {\cal H}~=~ {\cal H}_L~+~ {\cal H}_R$. Any function
$ F~\in ~{\cal A}$ and the Dirac operator $D$ acting on ${\cal H}$ are
represented by the $2\times 2$ matrices
\begin{equation}
 F~=~\pmatrix{f_1(x)&0\cr
                0&f_2(x)\cr}~~,~~
 D~=~ d.1 ~+~ D_Q ~=~\pmatrix{d & m \theta \cr
                   m{\bar \theta }& d\cr}
\end{equation}
where $m$ is a c-number with the dimension of mass and $\theta $ is a Clifford
element.

Althought our algebra $ {\cal A}~ =~ C^\infty({\cal M}) \oplus
C^\infty({\cal M}) $ is commutative, the geometry is not, since the
operator $ D$ contains a self-adjoint outerautomorphism part $ D_Q $.
Here we choose $ D_Q~=~D_Z~+~D_{\bar Z}~=DZ~\partial_z~+~D\bar Z~
\partial_{\bar
z}$. The derivatives and differential elements of our geometry are given as
follows
\begin{eqnarray}
D_\mu ~~~ = ~~ \pmatrix{\partial_\mu & 0\cr
                           0 & \partial_\mu\cr}
&~~,~~& DX^\mu ~~= ~~ \pmatrix{dx^\mu & 0\cr
                         0 & dx^\mu\cr} \nonumber\\
\partial_z ~ =~~ \pmatrix{0 & m\cr
                          0 & 0\cr}
&~~ ,~~ & DZ~=~\pmatrix{\theta & 0\cr
                 0& \theta\cr}   \nonumber \\
\partial_{\bar z} ~ = ~\pmatrix{0& 0\cr
                               m & 0\cr}
&~~ ,~~ & D{\bar Z} ~ = ~\pmatrix{{\bar \theta }& 0\cr
                                     0 & {\bar \theta }\cr}
\end{eqnarray}
The wedge product of any pair of differential elements is antisymmetric as in
the ordinary Riemannian geometry. It is straightforward, then, to construct
geometric objects following the sample of the ordinary geometry step by step.

\vglue 0.6cm

{\bf \noindent 3. The Standard Model}
\vglue 0.2cm
{\it \noindent 3.1 The fermionic sector and the gauge connection}
\vglue 0.1cm
\baselineskip=14pt
In our NC space-time gauge theories can be constructed in a complete analogue
of the ordinary one. It is not necessary to introduce any odd form to generate
the Higgs potential. The covariant derivative is : ${\cal D}= D + {\cal B}$.
By adjusting the fermionic sector, let us fix the gauge connection 1-form
${\cal B}$. The Lagrangian for the fermionic sector is as follows
\begin{equation}
{\cal L}_F =<\Psi ~|~ D + {\cal B} ~|~ \Psi >
\end{equation}
In the fermion sector we use the Dirac representation of the differential
elements $ dx^\mu =  \gamma^\mu $ and $\theta = \gamma^5 $. The scalar product
in Eq.~(3) is the Clifford trace. In order to have different Higgs
couplings to different types of particle we introduce the $3N_F\times 3N_F$
mixing matrix $ M $, where $N_F$ is the number of generations, in
the 1-form ${\cal B}$ as follows
 \begin{equation}
{\cal B} = \pmatrix{ gb_1 + 1/2.g'b_2 & \theta h\otimes M\cr
            \bar \theta  \bar h\otimes M^+ & g' b_2\cr}
 = DX^{\mu}~B_{\mu}+ \bar H \otimes M^+~D\bar Z + DZ ~H\otimes M
\end{equation}
 The gauge field and Higgs field matrices are given as follows:
\begin{equation}
B_{\mu}=\pmatrix{gb_{1\mu}+1/2.g'b_{2\mu}& 0\cr
           0& g' b_{2\mu}\cr} ~~~,~~~
H =~ \pmatrix{0&h\cr
         0&0\cr}    ~~~,~~~  \bar H~ =\pmatrix{0 & 0\cr
                                               \bar h & 0\cr}
\end{equation}
where $b_{1\mu}$ and $b_{2\mu}$ are respectively $ SU(2)$ and $ U(1)$ gauge
connections in the Standard Model. The Higgs fields $h$ and $\bar h$ are
automatically in doublets.
\vglue 0.4cm
{\it \noindent 3.2 The gauge-Higgs sector}
\vglue 0.1cm
The field strength is a direct generalization of the ordinary one
\begin{equation}
\Omega = D{\cal B} + {\cal B} \wedge {\cal B},
\end{equation}
\begin{eqnarray}
 \Omega_{\mu \nu} &=&{ 1\over 2}~(\partial_\mu B_\nu -\partial_\nu B_{\mu}
+[B_{\mu},B_{\nu}])= {1\over 2}~F_{\mu\nu}\nonumber\\
  \Omega_{\mu z}&=&{1\over 2}~(~(\partial_{\mu }H + B_\mu ~H-H~B_\mu)\otimes M
-\partial_zB_{\mu} )\nonumber\\
  \Omega_{\mu \bar z }&=&{1\over 2}~(~(\partial_{\mu }\bar H + B_\mu ~\bar H
-\bar H~ B_\mu)\otimes M^+ - \partial_{\bar z} B_{\mu } )\nonumber\\
\Omega_{z \bar z}&=&{1\over 2}~(\partial_{\bar z} H\otimes M-\partial_z \bar H
\otimes M^+ + \bar H  H\otimes M^+M - H \bar H \otimes MM^+)
\end{eqnarray}
Here we choose the following K\"ahler metric:
\begin{eqnarray}
<DX^\mu ~~DX^\nu > &=& ~~~g^{\mu \nu} \nonumber\\
<DX^\mu ~~DZ~ > &=& ~<DX^\mu ~~D\bar Z > ~=~ 0 \nonumber\\
<DZ ~~~D\bar Z~ > &=& <D\bar Z~~DZ > ~=~ 1
\end{eqnarray}
The Lagrangian of the pure Yang-Mills-Higgs sector now is the generalization
of the usual gauge Lagrangian $ {1\over g^2} <F^2>$
\begin{equation}
 {\cal L}_{YMH}= {1\over 3N_F} Tr G^{-2} < \Omega^2>
\end{equation}
In order to obtain the right kinetic terms for the gauge fields, we choose
the matrix G as
\begin{equation}
G~~=~~\pmatrix{ g & 0\cr
                0 & g' \sqrt{2/(2-(g'/g)^2)}\cr}
\end{equation}
The factor $ 1/3N_F$ is to cancel the trace of the $3N_F\times 3N_F $ unit
matrix. After shifting $ h \longrightarrow h -m\otimes M^{-1} $, we want to
obtain the Lagrangian for the gauge-Higgs sector of the Standard Model
\begin{equation}
{\cal L}_{YMH} = -{1\over 4} Tr F^2 + {1\over 2} {\cal D}_\mu \bar h
{\cal D}^\mu h + \lambda((\bar h h)^2 -2 v^2 {\bar h} h)+ constant
\end{equation}
In order to have the right terms in this Lagrangian, the coupling constants
$\lambda, g, g' $ and the mixing matrix $M$ must satisfy several constraints,
which can be reexpressed as the following mass formulas
\begin{eqnarray}
 m^2_h&~~=~~& 12 N_F M^2_W sin^2\theta_W/(2- sin^2\theta_W) \\
 M^2_H&~~=~~& 2 m^2_h \\
 m^2 &~~=~~& m^2_h
 \end{eqnarray}
where $m_h$ is the heaviest quark mass.
Assuming three generations, our mass formulas give the top quark mass
$m_t \approx  172 GeV$ and and the Higgs mass $ M_H \approx 241 Gev$ at the
electroweak scale. The recent CDF data [7] on the top quark mass
$m_t\sim 160-180 GeV$ strongly suggest that the energy scale of NC space-time
is the electroweak one.
\vglue 0.6cm
{\bf \noindent 4. Gravity}:
\vglue 0.4cm
In NC space-time, gravity also has new features. Next we introduce an
orthonormal basis of vielbein $\{E^A\}~ (A=a,z,\bar z)$. As a direct
generalization of vielbein,
$E^A\doteq DX^M E^A_M $ are 1-forms in NCG.
Here we are particularly interested in the self-adjoint vielbein
\begin{eqnarray}
E^a &\doteq& \pmatrix{e^a& 0\cr
                   0 & e^a\cr}  \nonumber\\
{}~~ \nonumber \\
E^z & \doteq & \pmatrix{\tilde g~a& m~\theta e^{-\kappa \sigma} \cr
                   0 & \tilde g ~a\cr}              \nonumber\\
{}~~\nonumber\\
E^{\bar z} & \doteq & \pmatrix{\tilde g~a& 0 \cr
                m~ \bar \theta e^{-\kappa \sigma} & \tilde g ~a\cr}
\end{eqnarray}
The vierbein $e^a$, the vector field $a$ and the scalar field $\sigma $ are
real.
The 'vielbein' $E^A$ reduces to the basis $DX^\mu, DZ$ and $D\bar Z$ if the
vector and scalar field vanish and the four-dimensional vierbein becomes flat.
Using the K\"ahler flat metric in Eq.~(8) for $z,\bar z $ and  $\eta ^{ab}$ for
$a,b=0,1,2,3$, we can redefine the vector field $a_\mu \rightarrow m~e^{-\kappa
\sigma}
 a_\mu$ and obtain
\begin{equation}\label{FINAL}
R=R_4 -2\kappa^2 \partial_\mu \sigma \partial^\mu \sigma -{1\over 4}
\tilde g^2 m^2 e^{-2\kappa \sigma}f_{\mu\nu}f^{\mu\nu}~,
\end{equation}
where $ f_{\mu\nu} = \partial_\mu a_\nu -  \partial_\nu a_\mu$
and $ R_4 $ is the 4-dimensional Ricci scalar.

The action is
\begin{equation}
S= {1\over 16\pi G} \int d^4x \sqrt{-det|g|} e^{-\kappa \sigma} R
\end{equation}

Fixing the right factors of kinetic terms, we have the following formulas
for the coupling constants $\tilde g$ and $\kappa $:\\
\begin{equation}
\tilde g = {\sqrt{16\pi G}\over m_h} ~~, ~~ \kappa = 2\sqrt{\pi G}
\end{equation}
where $ G $ is the Newton constant, hopefully not to be confused with the
gauge coupling matrix given previously.
It is interesting to notice that these couplings become much stronger
than the gravitational one at the scale of NC space-time.

\vglue 0.6cm
{\bf \noindent 4. Discussion \hfil}
\vglue 0.4cm
The predictions of the top quark and the Higgs masses are rather attractive.
They have a strong support from the recent experimental data on the top quark
mass. If the Higgs particle will be found near the
electroweak scale, it would definitely support the idea of NC space-time.
But, theorically, there are still some theoretical possibilities to be
clarified. It is possible, that the metric $< DZ ~D\bar Z>$ might depend on
the coupling matrix $G$ and/or the mixing matrix $M$. We have considered all
these possibilities and excluded them altogether as they all have wrong
physical implications. Here we do not
consider the unnatural possibilities, for example, that the quark and
the lepton sector might have different weight factors. The metric
$<DZ~ D\bar Z>$, in principle, can be any real number, however it can always be
absorbed into the mass scale $m$. The only alternative is the number of
generations $N_F$, which can be, of course, greater than $3$. As we expect that
the mass scale of the fourth generation is not as low as about $ 200 GeV$, the
mass formulas (12)-(14) are not valid at the electroweak scale. Given the
number of generations $N_F$, we can determine the heaviest quark mass $ m_h$ as
follows: The mass formula (12) is supposed to be valid at the scale $m=m_h$ of
NC space time, hence
\begin{equation}
m = \sqrt{12N_F} M_W(m)sin\theta_W(m)\sqrt{(2-sin^2\theta_W(m))}
\end{equation}
 As the right-hand side of this equation will evolve, using the values of $M_W$
and $ \theta_W$ at the electroweak scale as initial values we can determine
the intersection point of its curve with the line $ y= m$ and hence the scale
of NC space-time $m=m_h$. Having that scale we can use the mass formula (13) to
determine the Higgs mass.

If we will not find the Higgs particle at an energy scale as low as $241 GeV$
in
the near future, we really have to consider the possibilities of NC
space-time with the number of generations $N_F > 3 $. So, a confirmation of
NC space-time from the Standard Model is waiting for the observation of the
Higgs particles. Fortunately, the NC gravity can give an independent
prediction for the heaviest quark mass $m_h$ through Eq.~(18). We can couple
the NC gravity to the fermionic
sector of the Standard Model and obtain the following interaction terms
\begin{eqnarray}
{\cal L}_{f-a}&~~=~~&4\sqrt{\pi G}{\bar \Psi } a^\mu \gamma _\mu \gamma^5 \Psi
\\
{\cal L}_{f-\sigma}&~~=~~&-4m_h\sqrt{\pi G}{\bar \Psi }\gamma^5 \sigma \Psi
\end{eqnarray}
The couplings of the vector field $ a $ and the scalar field $\sigma $ to the
matter fields are universal and much stronger than the gravitational one.
So, their existence should be testable in cosmology.

Although, all these problems must be worked out in details, it is remarkable
that the idea of NC space-time can be certainly confirmed or denied by the
currently available facilities. Works to answer the remaining questions are
under progress.

\vglue 0.6cm
{\bf \noindent Acknowledgements \hfil}
\vglue 0.4cm
Thanks are due to Kameshwar C.Wali for the collaborative works and many helps.
I am grateful to Prof.A.Zichichi for a World Laboratory fellowship. I
would like also to thank Dr.Le Viet Dung for reading the manuscript.
\vglue 0.6cm

\vglue 0.5cm
\end{document}